\documentclass[10pt,conference]{IEEEtran}
\IEEEoverridecommandlockouts

\usepackage{hyperref}
\usepackage{cite}
\usepackage{amsmath,amssymb,amsfonts}
\usepackage{algorithmic}
\usepackage{graphicx}
\usepackage{textcomp}
\usepackage{xcolor}
\usepackage{algorithm}
\usepackage{etoolbox}
\usepackage[ruled,vlined,linesnumbered,algo2e]{algorithm2e}
\usepackage[linesnumbered,ruled,vlined]{algorithm2e} 
\usepackage{algorithmic}
\def\BibTeX{{\rm B\kern-.05em{\sc i\kern-.025em b}\kern-.08em
    T\kern-.1667em\lower.7ex\hbox{E}\kern-.125emX}}

\begin{document}
\title{ SNR and Resource Adaptive Deep JSCC for Distributed IoT Image Classification}
\author{
    \IEEEauthorblockN{
        Ali Waqas and 
        Sinem Coleri
    }
    \IEEEauthorblockA{Department of Electrical and Electronics Engineering, Koç University, Istanbul, Turkey}
    \IEEEauthorblockA{Emails: \{awaqas24, scoleri\}@ku.edu.tr}
    \thanks{ Sinem Coleri acknowledges the support of the Scientific and Technological Research Council of Turkey 2247-A National Leaders Research Grant \#121C314.}
}
\maketitle
\begin{abstract}

Sensor-based local inference at IoT devices faces severe computational limitations, often requiring data transmission over noisy wireless channels for server-side processing. To address this, split-network Deep Neural Network (DNN) based Joint Source-Channel Coding (JSCC) schemes are used to extract and transmit relevant features instead of raw data. However, most existing methods rely on fixed network splits and static configurations, lacking adaptability to varying computational budgets and channel conditions. In this paper, we propose a novel SNR- and computation-adaptive distributed CNN framework for wireless image classification across IoT devices and edge servers. We introduce a learning-assisted intelligent Genetic Algorithm (LAIGA) that efficiently explores the CNN hyperparameter space to optimize network configuration under given FLOPs constraints and given SNR. LAIGA intelligently discards the infeasible network configurations that exceed computational budget at IoT device. It also benefits from the Random Forests based learning assistance to avoid a thorough exploration of hyperparameter space and to induce application specific bias in candidate optimal configurations. Experimental results demonstrate that the proposed framework outperforms fixed-split architectures and existing SNR-adaptive methods, especially under low SNR and limited computational resources. We achieve a 10\% increase in classification accuracy as compared to existing JSCC based SNR-adaptive multilayer framework at an SNR as low as -10dB across a range of available computational budget (1M to 70M FLOPs) at IoT device.

 \end{abstract}
\section{Introduction}\label{sec1}
Edge-based inference using CCTV cameras, environmental sensors, and autonomous vehicles is paving the way for intelligence-integrated living in the forthcoming 6G era \cite{1celik}. However, these IoT devices often lack the computational and memory resources required to perform complex inference tasks locally, necessitating the transmission of collected data to remote servers for processing \cite{2}. To reduce bandwidth usage and latency, it is preferable to transmit only the most task-relevant portions of the data rather than the complete raw dataset. Identifying and extracting this relevant part while ensuring accurate inference on the server side presents a significant challenge.

Deep learning-based Joint Source-Channel Coding (JSCC) has shown promising results in extracting meaningful representations from data collected by IoT devices \cite{3,4,5}. These methods map the feature space from the final layer of a deep neural network (DNN) into a latent vector that can be directly transmitted, eliminating the need for separate source and channel encoders. An autoencoder-based JSCC scheme for wireless image transmission in \cite{3} outperforms traditional separate coding methods, even with small latent dimensions. In \cite{4}, a deep JSCC approach is applied to image retrieval, improving end-to-end accuracy and reducing encoding overhead by compressing features into a low-dimensional vector. A Convolutional Neural Network (CNN)-based encoder-decoder architecture in \cite{5} combines latent feature extraction with domain knowledge to enhance task-specific accuracy. However, these approaches require running the full network before latent space generation, which remains computationally intensive for resource-limited IoT devices.

To address computational constraints, several works have combined DNN splitting with JSCC to reduce processing at the IoT device \cite{6,7,8}. In this approach, the DNN is partitioned after a few layers, and the intermediate latent features are transmitted to the server, where the remaining network is executed. \cite{6} introduces a bottleneck-based partitioning scheme by training multiple architectures at different split points and selecting the optimal one. \cite{7} uses Linear Integer Programming (LIP) to decide which layers to process locally and which to offload.  \cite{8} proposes a split architecture that transmits intermediate features, balancing communication cost and accuracy. However, these methods either rely on fixed split points or require training multiple models, which is costly for IoT devices. Selecting the optimal split point remains difficult under varying computational capacities and channel conditions. \cite{9} addresses this by training fixed-split architectures over multiple SNRs, while \cite{10} adapts the split point based on SNR and task urgency using layer-wise successive transmissions. Despite these advances, most works keep network architecture parameters fixed, overlooking their significant impact on the computational load. These parameters should be optimized jointly with split point selection based on current channel conditions and the computational budget of the device.

\begin{figure*}[h]
    \centering
    \includegraphics[width=\textwidth]{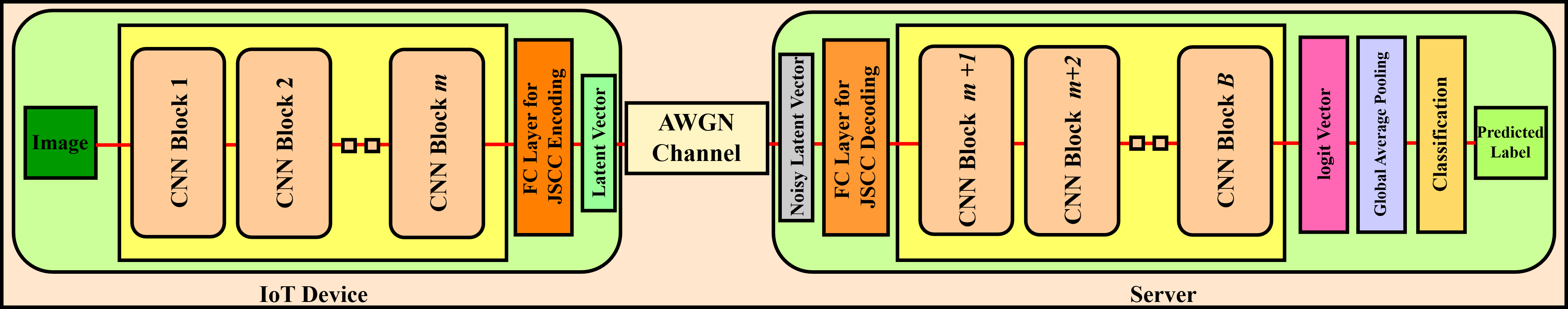}
    \vspace{-20pt}
    \caption{Block Diagram of the Proposed JSCC based Distributed Framework for Wireless Image Classification }
    \label{blockdiagram}
\end{figure*}

In this paper, we propose a JSCC-based adaptive wireless image classification framework that optimally selects distributed CNN parameters—including the network split point, number of filters, kernel size, and latent space dimension—based on the computational budget of the IoT device and the given channel SNR. The parameter optimization is driven by our proposed learning-assisted Intelligent Genetic Algorithm (LAIGA) that eliminates infeasible configurations early and evolves only valid candidate solutions, reducing both computation and latency compared to randomly searching through the parameter space. A set of Random Forest regressors, trained on an offline simulated dataset, provides application-specific estimates of accuracy and computational cost to guide the GA. While the framework is applied to image classification, it is generalizable to other tasks. To the best of our knowledge, this is the first work to address such comprehensive and dynamic optimization of CNN architecture and split configuration under strict computation and SNR constraints in IoT-based wireless image classification.

The remainder of the paper is organized as follows: Section II describes the system model. Section III details the proposed learning-assisted intelligent Genetic Algorithm. Section IV presents the performance evaluation, and Section V concludes the paper with final remarks.

\section{System Model}\label{sec2}

We consider a single IoT device that captures an image of an object requiring classification or recognition. Due to limited computational resources, the device cannot perform the task locally. To address this, we design a CNN architecture that can be adaptively split between the IoT device and a remote server, as illustrated in Fig. \ref{blockdiagram}. A JSCC-based scheme is applied at the split point to encode the feature space into a latent representation at the device and decode it back at the server.

The proposed network comprises a total of \( B \) CNN blocks and is optimally split after \( m \) blocks. Let the input image be denoted by \( \mathbf{X} \in \mathbb{R}^{H \times W \times C} \), where each pixel value is normalized to the range \([-1, 1]\) to prevent vanishing gradients. The normalized image \( \mathbf{X}_{\text{norm}} \) is processed through the first \( m \) CNN blocks located at the edge device. Each convolutional block performs the following operation:
\begin{equation}
    \mathbf{F}_l = f_{\theta_l}(\mathbf{F}_{l-1})
\end{equation}
where \( f_{\theta_l} \) denotes the convolution at layer \( l \), \( \theta_l \) represents the learnable parameters, \( \mathbf{F}_{l-1} \) is the input feature map, and \( \mathbf{F}_l \) is the output feature map.

The detailed operation within each block consists of a 2D convolution, batch normalization (BN), and a PReLU activation function \( \sigma(\cdot) \), expressed as:
\begin{equation}
    \mathbf{F}_l = \sigma\left(\text{BN}(\mathbf{W}_l * \mathbf{F}_{l-1} + \mathbf{b}_l)\right)
\end{equation}
where \( \mathbf{W}_l \) and \( \mathbf{b}_l \) denote the convolutional weights and bias at layer \( l \), respectively. The PReLU activation function is defined as:
\begin{equation}
    \sigma(x) = \max(0, x) + \alpha \min(0, x)
\end{equation}
with \( \alpha \) being a learnable parameter. After the \( m \)th block, the resulting feature map is projected onto a latent vector using JSCC encoding for transmission to the server.
 \begin{equation}
    \mathbf{h} = \mathbf{W}_{\text{latent}} \mathbf{F}_m + \mathbf{b}_{\text{latent}}, \quad \mathbf{h} \in \mathcal{C}^{S}
\end{equation}
where \( \mathbf{h} \) is the latent space vector, \( \mathbf{F}_m \) is the feature map at the \( m \)th layer, and \( \mathbf{W}_{\text{latent}} \), \( \mathbf{b}_{\text{latent}} \) are the weight matrix, bias of the fully connected (FC) layer used for projection, and \( S \) denotes the dimension of the latent vector.

This latent representation is transmitted over an Additive White Gaussian Noise (AWGN) channel. The received noisy latent vector \( \mathbf{y} \) at the server is given by:
\begin{equation}
    \mathbf{y} = \mathbf{h} + \mathbf{z}
\end{equation}
where \( \mathbf{z} \sim \mathcal{CN}(0, \sigma^2) \) is complex Gaussian noise. The noise variance \( \sigma^2 \) is computed from the SNR as:
\begin{equation}
    \sigma^2 = \frac{\mathbb{E}[\|\mathbf{h}\|^2]}{10^{\text{SNR(dB)} / 10}}
\end{equation}

At the server, the latent vector is decoded back to feature space using:
\begin{equation}
    \mathbf{\hat{F}}_m = \mathbf{W}_{\text{recover}} \mathbf{y} + \mathbf{b}_{\text{recover}}
\end{equation}
where \( \mathbf{\hat{F}}_m \) is the recovered feature map, and \( \mathbf{W}_{\text{recover}} \), \( \mathbf{b}_{\text{recover}} \) are the the weight matrix and bias of  the FC layer used for reconstruction. The remaining \( B - m \) CNN blocks are executed at the server. The final feature map is projected into a 1D logit vector with dimension equal to the number of output classes. A global average pooling layer is applied, and a prediction is made for the input image. This concludes the forward pass.

The predicted label is compared with the ground truth to compute the loss. For the backpropagation, gradients of all learnable parameters are calculated, and model weights are updated using the Adam optimizer.
\section{Learning-assisted Intelligent Genetic Algorithm}\label{sec3}

Our proposed intelligent optimization algorithm dynamically selects the CNN split point, number of output filters, kernel size, and latent space vector size to distribute the computational load between the IoT device and the server, based on the available compute capacity and channel SNR. Exhaustively evaluating all possible configurations is computationally infeasible. Furthermore, configurations that work well for one application may not generalize to others \cite{11,12}, highlighting the need for application-specific bias in the optimization process. To address this, we offline simulate a selected subset of CNN configurations from the parameter search space, tailored to the target application. This produces a domain-specific dataset that includes the computational cost (FLOPs) and classification accuracy for each SNR and candidate configuration. This dataset is used to train regression models that predict performance and resource requirements for unseen configurations. We use Random Forest Regression due to its generalization ability and resilience to overfitting across a wide parameter space. By guiding the search toward high-accuracy, resource-feasible configurations, this learning-based approach accelerates convergence through the large design space.

\subsection{Optimization Problem}

We aim to maximize image classification accuracy under a given computational budget and channel SNR. The optimization problem is formulated as:
\begin{equation}
    \max_{\Theta} A(\Theta)
\end{equation}
subject to:
\begin{equation}
    F(\Theta) \leq F_{\text{max}}
\end{equation}
\begin{equation}
    \text{SNR} = q
\end{equation}
where \( \Theta = \{ f, k, l_s, m \} \) denotes the set of hyperparameters, \( f \) is the number of filters, \( k \) is the kernel size, \( l_s \) is the latent space dimension, and \( m \) is the number of CNN layers executed at the IoT device (i.e., the split point). \( A(\Theta) \) represents the classification accuracy, and \( F(\Theta) \) is the computational load of the configuration. The constraint \( F_{\text{max}} \) defines the available computational budget, and \( q \) is the current channel SNR.

The computational complexity for image classification is determined by the CNN parameters. For the \( i \)th convolutional layer, the FLOPs are given by:
\begin{equation}
    F_i = 2 \times \left(k_i^2 \times f_{\text{in}_i} \times f_{\text{out}_i} \times h_i \times w_i\right)
\end{equation}
where \( k_i \) is the kernel size, \( f_{\text{in}_i} \) and \( f_{\text{out}_i} \) are the number of input and output filters, and $h_i$ and $w_i$ are the height and width of the feature map at layer \( i \), respectively. For the fully connected layer mapping the feature space of \(mth\) convolutional layer to the latent space, the number of FLOPs is given by:
\begin{equation}
    F_{\text{latent}} = 2 \times \left(f_{\text{out}_m} \times h_m \times w_m \times l_s\right)
\end{equation}
The total computational cost at the IoT device is:
\begin{equation}
    F(\Theta) = \sum_{i=1}^{m} F_i + F_{\text{latent}}
\end{equation}

Increasing the number of layers \( m \) at the IoT device raises computational cost but allows for deeper feature extraction. When resources permit, assigning more layers to the edge improves task-relevant accuracy at the server.


\subsection{Random Forest Regressors}

Due to the high computational cost of full grid search, we construct the dataset \( \mathcal{D} \) by selectively sampling representative configurations from the parameter space and recording their accuracies and FLOPs at selected SNR levels. These offline simulations cover only a manageable subset of the full space. The resulting dataset is used to train Random Forest regressors, which predict performance metrics for unseen configurations, thereby accelerating the optimization process.

Random Forests operate in an ensemble learning fashion by building multiple decision trees using bootstrap sampling. The dataset \( \mathcal{D} \) is split into multiple subsets with replacement, each used to train a separate tree. For a new configuration, the predicted FLOPs and accuracy are obtained by averaging the outputs of all trees, reducing the risk of overfitting. This ensemble approach provides strong generalization across the full parameter space and is well-suited to our application \cite{13}.

 We train two separate models: (i) a FLOPs regressor \( R_F \) that estimates computational cost \( \hat{F}(\Theta) \), and (ii) an accuracy regressor \( R_A \) that estimates task performance \( \hat{A}(\Theta) \) for given configuration \( \Theta \) at a specified SNR. This approach not only reduces the number of required offline simulations but also incorporates application-specific bias from real data.

\begin{equation}
    \hat{F}(\Theta) = R_F(f, k, l_s, m)
\end{equation}
\begin{equation}
    \hat{A}(\Theta) = R_A(f, k, l_s, m)
\end{equation}

\subsection{Intelligent Genetic Algorithm}

We propose a learning-assisted intelligent Genetic Algorithm (LAIGA) described in Algorithm~1. The algorithm eliminates infeasible configurations early, enabling an efficient search for the optimal configuration \( \Theta_{\text{optimal}} \) under computational and SNR constraints. GA is well-suited for our discrete parameter space, unlike gradient-based methods (e.g., SGD, Adam), which require differentiable objective functions and are less effective in such scenarios \cite{14,15}.

The goal is to determine the optimal values of \( \Theta = \{f, k, l_s, m\} \) that maximize image classification accuracy while satisfying constraints (9) and (10). We initialize a population of \( V \) random configurations from the search space (Line $1$). For configurations already present in the offline dataset, FLOPs and accuracy values are directly retrieved (Lines $4-5$). Otherwise, they are predicted using the trained regressors \( R_F \) and \( R_A \) from Equations (14) and (15) (Lines $6-7$). Configurations violating the FLOPs constraint in (9) are discarded by assigning them a fitness score of \( -\infty \) (Lines $8-9$).
Valid configurations are evaluated using a fitness function based on accuracy and resource utilization (Lines $10-11$). Each configuration  $c$ receives a fitness tuple:  $\text{fitness}(c) = (\left (A(c) + U(c), 
F_{\text{max}}-F(c), \right )$ where  $U(c) = \max\left(0,\ \frac{F(c) - 0.9 \times F_{\text{max}}}{0.1 \times F_{\text{max}}}\right)$.

The utilization $U(c)$ rewards configurations that utilize 90--100\% of the available computational budget and penalizes under-utilization, while minimizing the FLOPs gap. 
\begin{flushright}
\begin{minipage}{0.95\columnwidth}
\begin{algorithm}[H]
\caption{Learning-Assisted Intelligent Genetic Algorithm (LAIGA)}
\KwIn{
FLOPs budget \( F_{\text{max}} \);\\
Channel SNR \( q \in \{q_1, q_2, \ldots, q_n\} \);\\
Offline dataset \( \mathcal{D}  \);\\
Trained regressors: \( R_F \), \( R_A \);\\
GA parameters: population size \( V \), generations \( G \), tournament size \( T \), crossover probability \( cxpb \), mutation probability \( mutpb \)
}
\KwOut{Optimal configuration \( \Theta_{\text{optimal}} \)}

\BlankLine
Initialize population \( P \) with \( V \) random configurations\;


\For{generation \( = 1 \) to \( G \)}{
  \ForEach{\( c \in P \)}{
    \eIf{\( c \in \mathcal{D} \)}{
      Retrieve $F(c)$ and $A(c)$ from \( \mathcal{D} \)\;
    }{
      \( F(c) \gets R_F(c) \), \( A(c) \gets R_A(c) \)\;
    }
    \eIf{\( F(c) > F_{\text{max}} \)}{
      \( \text{fitness}(c) \gets (-\infty, \infty) \)\;
    }{
      \( \text{fitness}(c) = (\left(A(c) + U(c), \\
 F_{\text{max}}-F(c) \right) \)\;
    }
  }
  Perform tournament selection of size \( T \)\;

    Apply crossover with probability \( cxpb \)\;

  Apply mutation with probability \( mutpb \)\;
  
  Update population \( P \) with new configurations\;
 }
Select configuration with highest accuracy from final population

\If{tie}{select configuration with minimum FLOPs gap }\;
\Return \( \Theta_{\text{optimal}} \)\;
\end{algorithm}
\end{minipage}
\end{flushright}

We apply a \textit{tournament selection} strategy to promote strong yet diverse candidates (Line $12$). Specifically, \( V \) overlapping subsets, each of size \( T \), are formed from the current population. The top-performing configuration in each subset, based on the fitness score, is selected as a \textit{parent configuration}. These parent configurations represent strong candidates from the current generation and are used to produce the next generation. Each pair of parents undergoes uniform crossover with probability \( cxpb \), exchanging parameter values to promote exploration of the search space (Line $13$). Offspring configurations are then subjected to mutation with probability \( mutpb \), where one parameter is randomly altered within its valid range (Line $14$). This process yields the next generation of configurations, which are evaluated and re-entered into the optimization cycle (Line $15$).

The GA is executed for \( G \) generations, iteratively refining the configurations of \( \Theta \) under the specified constraints. The population update rule at each generation is given by:
\begin{equation}
    P_{t+1} = \text{Selection} \left( \text{Mutation} \left( \text{Crossover}(P_t) \right) \right)
\end{equation}
where \( P_t \) is the population at generation \( t \), and \( P_{t+1} \) is the evolved population. 

From the final generation, we select the configuration with the highest predicted accuracy (Line $16$). In case of a tie, the configuration with the smallest FLOPs gap is chosen (Lines $17-18$). The resulting \( \Theta_{\text{optimal}} \) is then used to configure the network in Fig.~1, ensuring optimal image classification performance within the given computational and SNR constraints.

\section{Performance Evaluation}\label{sec5}

This section evaluates the performance of the proposed framework, LAIGA, for wireless image classification under computational constraints while providing SNR adaptivity. We compare LAIGA with two benchmark schemes: (i) the fixed split-point architecture from \cite{9}, and (ii) an SNR-adaptive JSCC scheme, ADJSCC-\textit{l}, which transmits base information with high compression in the first layer and refines quality in subsequent layers through additional transmissions \cite{10}.

All simulations are implemented in PyTorch and executed on an NVIDIA RTX A2000 GPU. Experiments are conducted using the CIFAR-10 dataset. For each configuration, we report training loss, accuracy per epoch, and associated FLOPs. Every model is trained for 30 epochs and evaluated over a range of SNR values. We consider a discrete parameter space defined as \(f \in \{8, 16, 32, 64, 128, 256\}\), \(k \in \{2, 3, 4, 5, 6, 7, 8, 9\}\),  \(l_s \in \{32, 64, 128, 256, 512\}\), and \(m \in \{1, 2, 3, 4, 5, 6\}\).

Instead of using a fixed number of filters across layers, we progressively increase the filter count. For example, if the chosen configuration sets \( f = 16 \) and \( m = 3 \), the output filters in the three edge layers will be 16, 32, and 64, respectively. This gradual expansion improves accuracy compared to fixed-filter schemes by widening the receptive field across layers \cite{16}. For generating the dataset \( \mathcal{D} \) used to train the Random Forest regressors, we iterate kernel sizes and latent space dimensions across layers for selected filter values. A total of 114 configurations are simulated, each trained for robustness by randomly assigning an SNR value from the set \(SNR \in \{-20, -15, -10, -5, 0, 5, 10, 15, 20, 25\}\) in dB.

Classification performance is evaluated using Cross-Entropy Loss, given by $\mathcal{L} = -\sum_{i=1}^{N} y_i \log(\hat{y}_i)$,
where \( N \) is the number of classes, \( y_i \) is the true label (one-hot encoded), and \( \hat{y}_i \) is the predicted probability for class \( i \).
We use label smoothing of 0.1 for \(\mathcal{L}\) and learning rate of 0.001 for the optimizer along with a batch size of 32. 

For the proposed LAIGA framework, we use standard parameter values: population size \( V = 20 \), number of generations \( G = 50 \), tournament size \( T = 3 \), uniform crossover probability \( cxpb = 0.5 \), and mutation probability \( mutpb = 0.2 \). To explore the search space thoroughly, we run 10 independent instances of the learning-assisted GA for each combination of FLOPs and SNR constraints and select the best resulting configuration.

\begin{figure}[ht]
    \centering
    \includegraphics[width=1\linewidth]{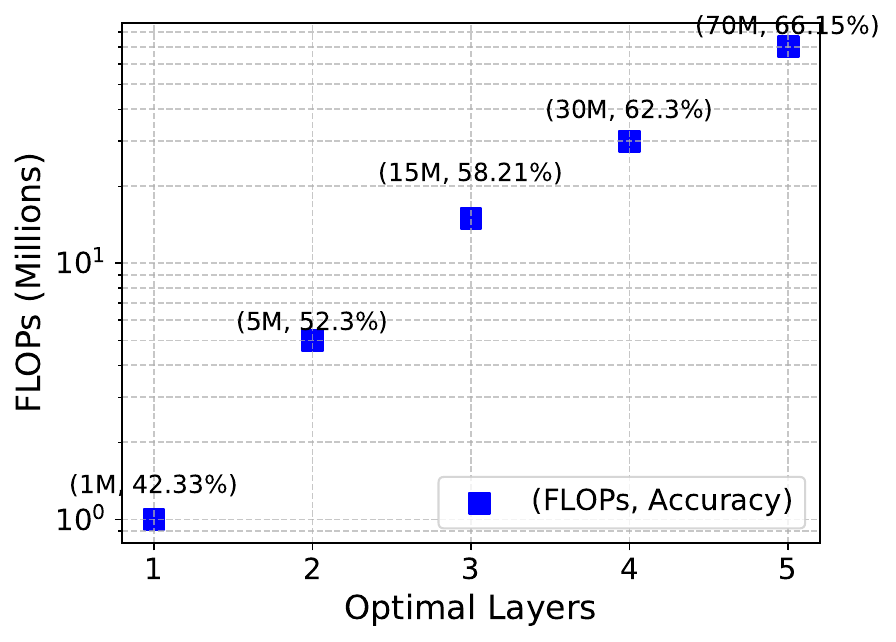} 
    \vspace{-20pt}
    \caption{Variation in required FLOPs with Number of Layers
    @ \textit{SNR} = -10dB, \textit{f} = 32, \textit{k} = 3, \textit{l\textsubscript{s}} = 128}
    \label{LF}
\end{figure}

Figure~\ref{LF} shows the variation of the computational demand with the number of layers executed at the IoT device, using a fixed SNR of \(-10\,\text{dB}\) and the corresponding optimal parameters. FLOPs are plotted on a logarithmic scale for better visual representation. As shown, the fixed-layer architecture from \cite{9} incurs a constant FLOPs cost, making it unsuitable for scenarios with limited computational resources. At a low SNR of \(-10\,\text{dB}\), the classification accuracy ranges from 42.33\% to 66.15\%, depending on the number of layers executed at the IoT device and the available computational budget.

\begin{figure}[ht]
    \centering
    \includegraphics[width=1\linewidth]{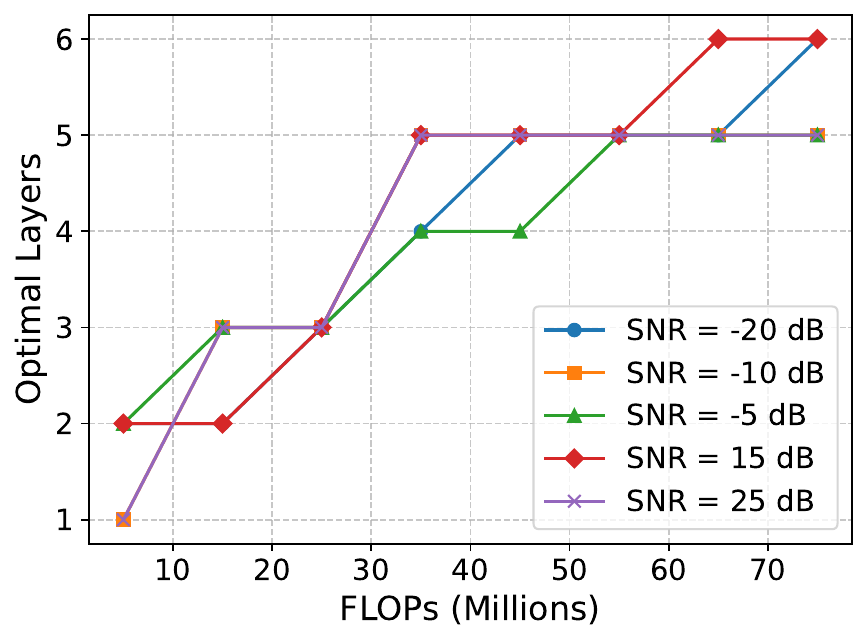} 
    \vspace{-20pt}
    \caption{Optimal number of layers for different FLOP budgets and channel SNRs.}
    \label{LF1}
\end{figure}

The behavior of individual parameters such as kernel size, number of filters, and latent space dimension does not exhibit a consistent pattern with changes in the available computational budget.
However, as shown in Fig.~\ref{LF1}, the number of selected layers at the IoT device follows a monotonically non-decreasing trend, regardless of the SNR values.

\begin{figure}[ht]
    \centering
    \includegraphics[width=1\linewidth]{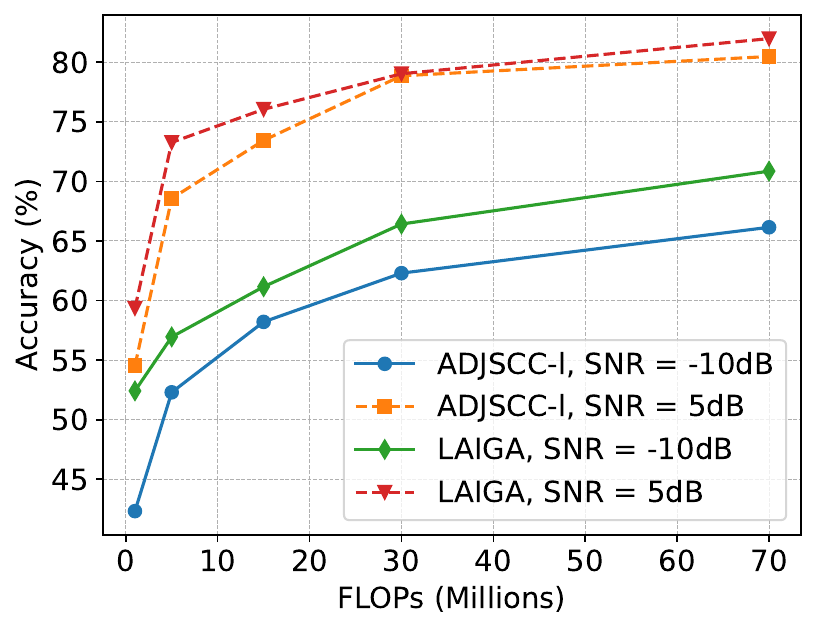} 
    \vspace{-20pt}
    \caption{Accuracy performance as a function of  FLOP budget for different channel SNRs.}
    \label{AF}
\end{figure}

Figure~\ref{AF} presents the accuracy of the proposed framework LAIGA compared to ADJSCC-\textit{l}, evaluated across varying computational budgets and fixed SNR values. LAIGA consistently outperforms ADJSCC-\textit{l} across all budget levels, primarily due to its computationally adaptive design. Even under low computational budgets, LAIGA enables full feature space propagation through the combined CNN layers on both the IoT device and the server. In contrast, ADJSCC-\textit{l} relies on layer-wise transmissions only when refinement is required. The performance gain is especially notable at low SNR (\(-10\,\text{dB}\)) and tight computational budgets (1M–10M FLOPs), where our method achieves up to a 10\% accuracy improvement. This gain is attributed to the SNR-robust training of the supervising dataset used during model optimization. 

\begin{figure}[ht]
    \centering
    \includegraphics[width=1\linewidth]{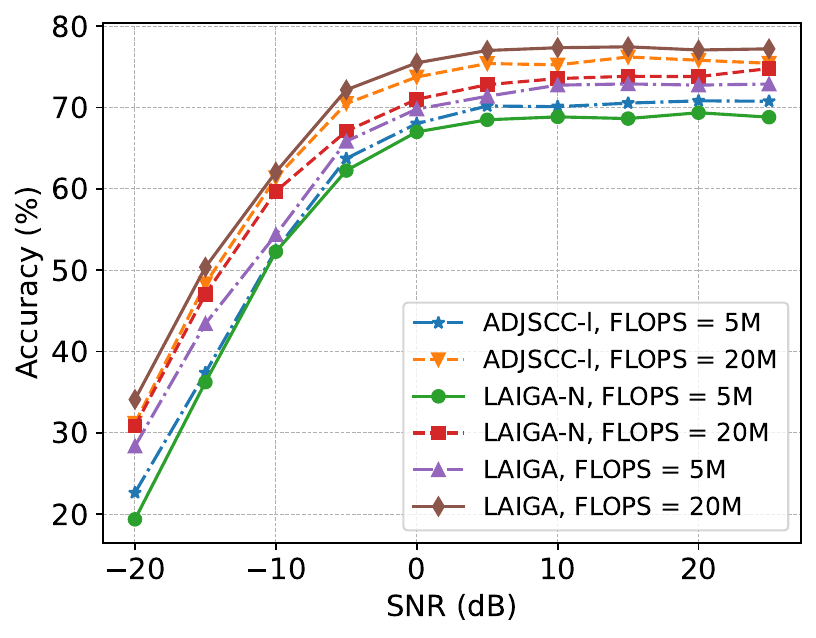} 
    \vspace{-20pt}
    \caption{Accuracy performance as a function of  channel SNR for different FLOP budgets. }
    \vspace{-1em}
    \label{SA}
\end{figure}

Figure~\ref{SA} shows the accuracy of the proposed framework LAIGA compared to ADJSCC-\textit{l} and the SNR non-adaptive variant of our method--LAIGA-N, evaluated across a range of SNRs under fixed computational constraints. LAIGA consistently outperforms both baselines. While ADJSCC-\textit{l} performs well at higher SNRs, its accuracy at lower SNRs is comparable to LAIGA-N. Unlike ADJSCC-\textit{l}, which adapts only to SNR, our framework jointly considers both computational and SNR constraints, yielding better performance under limited resources. Additionally, our SNR-robust training strategy enables strong generalization across the entire SNR range—even when explicit SNR adaptivity is not applied. This generalization makes the non-adaptive version of our framework comparable to the SNR-adaptive ADJSCC-\textit{l}, particularly at low SNRs. These results underscore the importance of SNR-robust training and adaptive optimization in enhancing classification performance over noisy wireless channels, especially for resource-constrained IoT devices.

\section{Conclusion}

We propose a deep JSCC-based distributed wireless image classification framework optimized for both computational and SNR adaptivity. The framework maximizes classification accuracy under strict computational constraints at the IoT device for a given channel SNR. An adaptively configurable CNN is optimized using a learning-assisted intelligent Genetic Algorithm guided by Random Forest regressors. This approach introduces application-specific bias while significantly reducing the need for extensive offline simulations. Our adaptive-split architecture demonstrates greater flexibility in resource allocation compared to the fixed split-point baseline. It also outperforms prior methods focused solely on layer optimization, particularly under low SNR and tight FLOPs budget conditions common in IoT scenarios. Finally, we show that jointly adapting to SNR and computational constraints yields better performance than computation-aware optimization alone. In future work, we plan to investigate optimal configurations for applications such as image, speech, and video classification to identify explainable factors behind application-specific biases.

\bibliographystyle{IEEEtran}\bibliography{bib_ali}
\end{document}